\documentclass[ aps, pra, twocolumn, showpacs,nofootinbib,superscriptaddress,longbibliography]{revtex4-2}

\usepackage{graphicx}
\usepackage{hyperref}
\usepackage{dcolumn}
\usepackage{bm}
\usepackage{amsmath}
\usepackage{bbold}
\usepackage{mathtools}
\usepackage{amsmath, bigstrut}
\usepackage{nicematrix, tikz}
\usetikzlibrary{fit}
\usetikzlibrary{matrix,decorations.pathreplacing}
\usepackage{amssymb}
\usepackage{mathrsfs}
\usepackage{natbib}
\usepackage{notoccite}

\usepackage{graphicx}

\begin{document}

\preprint{APS/123-QED}

\title{Qualitative differences in the robust controllability of model two-qubit systems}

\author{Anirban Dey}
\affiliation{School of Mathematical and Physical Sciences, Macquarie University,  NSW 2109, Australia}
\affiliation{ARC Centre of Excellence for Engineered Quantum Systems, Macquarie University,  NSW 2109, Australia}
\author{Mattias T. Johnsson}
\affiliation{School of Mathematical and Physical Sciences, Macquarie University,  NSW 2109, Australia}
\author{Daniel Burgarth}
\affiliation{School of Mathematical and Physical Sciences, Macquarie University,  NSW 2109, Australia}
\affiliation{Physics Department, Friedrich-Alexander Universit\"{a}t of Erlangen-Nuremberg, Staudtstr. 7, 91058 Erlangen, Germany}

\date{\today}

\begin{abstract}
The precise implementation and manipulation of quantum gates is key to extracting advantages from future quantum technologies. Achieving this requires very accurate control over the quantum system. If one has complete knowledge about a Hamiltonian, accurate manipulation of the system is possible. However, in real scenarios, there will often be some uncertainty in the parameters of the Hamiltonian, which makes full control of the system either difficult or impossible. In this paper we consider two model Hamiltonians with a continuous parameter that is partly unknown. We assess robust controllability against this parameter uncertainty using existing theoretical frameworks and take a numerical route by discretizing the unknown parameter in the cases where we cannot predict controllability. Furthermore, we introduce a penalty term into the fidelity function to optimize control pulses, enhancing robustness against the influence of parameter fluctuations. Within our framework, we analyze the qualitative differences in the robust controllability of the two systems.

\end{abstract}

\maketitle


\section{Introduction}
Future quantum technologies and quantum engineering aim to harness the potential of quantum mechanics, outperforming the best possible classical computers and other existing technologies such as metrology and sensing. Building such quantum technologies requires precise quantum states as well as accurate manipulation of those states; for example, one needs high-quality qubits and high-quality gates to implement quantum computation. Realization of gates with very high fidelity is difficult to achieve due to quantum systems being easily perturbed and prone to error. The mitigation of these errors and successful implementation of error-resilient gates through Hamiltonian dynamics is the province of quantum control, a major focus of ongoing research efforts \cite{QcomNISQ, WEIDNERrobust, WerninghausleakageNPJ,KochquantumEPJ, PhysRevLett.132.193801, KrantzAPR, StefanatosEPL,DongdoublyPRX}. \\
The study of quantum control enables us to explore the reachable set of implementable unitary gates through the time dynamics of a given Hamiltonian with some time-controllable parts. We can often express this type of Hamiltonian as $H(t) = H_d + \sum_{j=1}^m f_j(t) H_j$, where $H_d$ is the drift Hamiltonian, which we cannot control, $H_j$ are the control Hamiltonians, and $f_j(t)$ are the time-dependent control amplitudes. The unitary evolution then follows as $U(t) = \mathcal{T} e^{-i \int_0^t H(\tau) d\tau}$, where $\mathcal{T}$ is the time-ordering operator.\\
A fully controllable system would allow for the implementation of any unitary gate in a finite time through its Hamiltonian dynamics. For a finite-dimensional system, one can establish the full controllability of a system using the Lie-algebra rank criterion (LARC). In the presence of a drift, the LARC implicitly relies on the quantum recurrence theorem \cite{BocchieriQRT, SchulmanNote, ryosukePRArobust}, which will be relevant later. It states that for a Hamiltonian H, for any arbitrary error $ \varepsilon >0$, and any time duration $T >0$, there exists a time instance $T_{r} \geq T$ such that $\big \Vert \mathbb{1}-e^{-iHT_{r}}\big \Vert <\varepsilon$. This implies that the system will return to its initial state arbitrarily close at the recurrence time $T_r$. The theorem enables us to effectively implement the dynamics induced by the negative of the drift Hamiltonian.\\
There are two key formulas which are used to determine the LARC condition. These formulas are defined for any matrices A and B, with t$\in \mathbb{R}$, as follows:\\
(i) Trotter expansion formula: 
\begin{equation}\label{trotter}
\lim_{n \to \infty} \Big(e^{\frac{At}{n}} e^{\frac{Bt}{n}}\Big)^{n} = e^{(A+B)t},
\end{equation}
(ii) Commutator expansion formula: 
\begin{equation}\label{commutator}
\lim_{n\to \infty} \Big(e^{-\frac{A t}{n}}e^{-\frac{B t}{n}} e^{\frac{A t}{n}}e^{\frac{B t}{n}}\Big)^{n^{2}} = e^{[A, B]t^{2}}.
\end{equation}
These will play an important role later. In the practice of quantum control, it is very difficult to achieve precise manipulation of a quantum system. A very common source of error in real systems arises from the fact that one or more parameters in the system are unknown, or known only within some range, such as the presence of a stray magnetic field, instability in the control magnitude, the exact intensity of the laser manipulating the system or the orientation of a N-V crystal symmetry axis. The situation becomes even more challenging if the control parameters of the hardware drift over time. We need the implementation of robust gates against such errors as well as systematic calibration of the qubits to avoid errors due to drifting hardware parameters \cite{RealMajumderNPJ, YuvalPRXexperimental, Perf.WHITEprapp, UniversalRiesIEE}. Several research efforts are actively going into developing error-robust pulse sequence for manipulating quantum systems. Robustness means, these pulse sequence can function reliably despite parameter variations. This enables desired operation up to required accuracy in practical systems such as superconducting qubits in IBM \cite{harisonIOPQSTsoftware, YuvalPRXexperimental, ivanovPRLhigh, Perf.WHITEprapp, ZeyuanARXIVquantum, KaitlinproFIP, semolaIEEEdeep} and Google quantum computers \cite{AcharyaNATUREsupressing, frankARXIVobservation, nicolasPRAPPintegrated, KellyPRAscalable, julianARXIVphysical, cupjinPRLquantum, linglingACMdesigning, yilunACMautomatic} or trapped-ion qubit systems \cite{jakePRRtrapped, bermudezPRArobust, yotamPRLrobust, yotamPRLeentangle, valahuJPBquantum}.

In most situations, the parameters can take continuous values, which essentially transforms the control problem from finite dimension to an infinite dimension control, significantly increasing its complexity. Actually, shown by \cite{karineCMPcontrollability}, the best hope in this scenario is approximate robust controllability. In this paper, we thereafter only consider approximate controllability. The controllability of such systems is well understod when the values of the unknown parameters are given in either in a finite set or a discretized subset of a given continuous set, with useful lemmas establishing criteria for full controllability \cite{allgcontrol, mohammedEJCensemble, gabrielJPAMGoptimal, dirrMCSSaccess, indraIEEEcontrollability}. Deciding controllability for a continuous unknown parameter, however, remains challenging. In particular, it is unclear whether the robust controllability of the systems with unknown parameters in all discretized approximations implies robust controllability of the continuous one \cite{dirrMCSSaccess}. Therefore, the study of controllability of such systems is very interesting. 

A prototypical example of such a system is $H(\omega, t) = \omega H_d + f(t)H_c$ where $\omega$ is an unknown parameter in a range $[\omega_{0}, \omega_{1}]$. To analyze controllability, one might employ a Lie-algebraic approach along with polynomial approximation techniques, initiated by Li and Khaneja \cite{liIEEEensemble, liPRAcontrol}. While these methods have been successful in analyzing controllability, their applicability is often limited to simplistic systems \cite{ryosukePRArobust, liIEEEensemble, liPRAcontrol, karineCMPcontrollability, mohamedEJCensemble}. In our study, we explore two two-qubit systems,
\begin{equation}\label{system1}
H_d = \omega X \otimes I + X\otimes X + Y\otimes Y + Z\otimes Z,  H_{c} = Z\otimes I,
\end{equation} 
and
\begin{equation}\label{system}
H_d = \omega (X\otimes I + I \otimes X ) + Z \otimes I + Y \otimes Y + Z \otimes Z, H_{c} = X \otimes I,
\end{equation}
where $X, Y$ and $Z$ are the Pauli operators and $\omega$ is the unknown parameter. These two systems represent practical scenarios for real quantum systems. Our aim is to investigate their robust controllability while comparing their relative ease of implementation, as robust controllability reveals the feasibility of their practical realization.

While the first system is robustly controllable for a continuous and compact subset of the unknown parameter $\omega$~\cite{ryosukePRArobust}, despite its apparent simplicity, conventional schemes~\cite{ryosukePRArobust, liIEEEensemble, liPRAcontrol, karineCMPcontrollability, mohamedEJCensemble} have not been able to conclusively determine the robust controllability of the later one. In such a scenario, a numerical approach can provide a viable solution~\cite{KochquantumEPJ,OptimGe}. In the numerical approach, one discretizes the unknown parameter $\omega$ within a range $[\omega_0, \omega_1]$ and seeks an optimal pulse that works for each discrete value of $\omega$~\cite{ryosukePRArobust}. We employ gradient-based optimization techniques to search for an optimal control pulse. However, achieving robust control typically comes at the cost of longer pulse durations. Therefore, we examine the controllability of the two systems with varying control times.  Additionally, we design optimized pulses using a modified fidelity measure that incorporates a gradient-dependent penalty term. This penalty term is constructed from the weighted gradient of the fidelity with respect to the unknown parameter. To demonstrate our numerical approach, we investigate the robust controllability of the system described by Eq. [\ref{system1}] and [\ref{system}], and qualitatively compared their controllability. 	

The paper is organized as follows. We consider two-qubit systems with an unknown parameter that takes continuous values and discuss its robust controllability using theoretical techniques in Sec. \ref{RC}. In Sec. \ref{NA}, we move towards a numerical approach by discretizing the unknown parameter and investigating robust controllability. We discuss our numerical mthods, present our findings, and explore robust controllability by relating the continuous unknown parameter in a closed interval, while comparing the controllability of two different systems. In Sec. \ref{MF}, we introduce a modified fidelity by introducing a penalty term in our numerical optimization and assess impact on the robustness of control pulses, further delving deeper into the comparison between the two systems. Finally, we summarize our findings in Sec. \ref{SD}.

\section{Robust controllability} \label{RC}
In this section, we employ polynomial approximation to assess robust controllability. To illustrate, let us consider a single qubit problem of the form $H(t) = H_{d} + f(t) H_{c}$ with only a single control \cite{ryosukePRArobust, liIEEEensemble}. Let the system be $H_{d}(\omega) = \omega X, H_{c} = Z$ where $\omega$ is an unknown parameter within the interval $\omega \in [\omega_{0}, \omega_{1}] \subset \mathbb{R^{+}}$. We use the notation  $\mathbb{L}_{H(t)}$ to denote all the Hamiltonians whose time evolution can be effectively (approximately) generated, referred as simulable Hamiltonians. 

First, we apply a bang-bang style $\delta$-pulse to implement $Z H_{d}(\omega) Z = -H_{d}(\omega) \in \mathbb{L}_{H(t)}$. Using the commutator approximation formula, we obtain $[iH_{d}(\omega), iH_{c}] = \omega Y \in \mathbb{L}_{H(t)}$ and $[i\omega Y, iH_{d}(\omega)] = \omega^{2}H_{c} \in \mathbb{L}_{H(t)}$. Repeated application of commutators yields $\omega^{2n + 1} X, \omega^{2n + 1} Y \in \mathbb{L}_{H(t)}$ for $n \in \mathbb{N}$. The Trotter formula then further allows to simulate $\omega P_{1}(\omega^{2})X $, where $P_{1}(\omega^{2})$ is even polynomial functions of $\omega$. 

Now, we can extend the function $\frac{C_{1}}{\omega}$ on $[\omega_0, \omega_1]$ to an even function with an arbitrary small error. Consequently, we obtain $\omega P_{1}(\omega^{2}) = C_{1}$, where $C_1\in\mathbb{R}$ is a constant. This will allow any arbitrary rotation around X and Y, leading us to perform any quantum gate robustly independent of $\omega$ using this system. 

Next, we apply the polynomial approximation to two-qubit Hamiltonians presented in Eq. [\ref{system1}] and [\ref{system}] of the form $H(t) = H_{d}(\omega) + f(t)H_{c}$.\\~\\
{\bf System A:} $H_d(\omega) = \omega X \otimes I + X\otimes X + Y\otimes Y + Z\otimes Z$ and $H_{c} = Z\otimes I$ (see ref. \cite{ryosukePRArobust}).\\~\\
We can demonstrate robust controllability through the following steps:\\~\\
I. We can simulate $H_{d}(\omega)$ by switching off the control and simulate $\pm H_{c}$ using a strong $f(t)$.\\~\\
II. Using the $\delta$ pulse technique $H_{c}\big(H_{d}(\omega)\big)H_{c}$, we get $-H_{d}(\omega)+ 2 Z \otimes Z$.\\~\\
III. Using the Trotter expansion formula we can then simulate $H_{d}(\omega) - H_{d}(\omega)+ 2 Z\otimes Z = 2 Z\otimes Z$. The quantum recurrence theorem will lead us to simulate $- Z\otimes Z$ as well.\\~\\
IV. Implementing $Z\otimes Z\big(H_d(\omega)\big)Z \otimes Z$ and using Trotterization with $\pm Z\otimes Z$ will simulate $-H_{d}(\omega)$. \\~\\
V. Using the commutator \( [Z \otimes Z, [H_d(\omega), Z \otimes Z]] \), we can  simulate \( \pm \omega X \otimes I \). Both \( \pm \) terms are needed to implement the commutator formula.\\~\\
VI. Trotterizing $H_d(\omega)$ with $\pm\omega X\otimes I$ will simulate $\pm (X \otimes X + Y\otimes Y + Z\otimes Z)$.\\~\\
The following sequence will lead us to $[H_{d}(\omega), Z\otimes Z] \propto -\omega Y\otimes Z \rightarrow [\omega X\otimes I, \omega Y \otimes Z]\propto \omega^{2} Z \otimes Z \rightarrow [\omega^{2} Z\otimes Z, [H_{d}(\omega), Z\otimes Z]] \propto \omega^{3} X\otimes I$. By repeating the sequence, we can achieve $\omega^{2n+1} X \otimes I$. Hence, for an interval $[\omega_{0}, \omega_{1}]$ with $\omega_{0}\omega_{1} > 0$ we can achieve robust controllability of the system A. \\~\\
 {\bf System B:} $H_d (\omega) = \omega (X\otimes X + I \otimes X ) + Z \otimes I + Y \otimes Y + Z \otimes Z$, $H_{c} = X \otimes I$.\\~\\
 Let us try the same procedure as for system A,\\
 I. By setting the control amplitude to zero, we get $H_{d}(\omega)$.\\~\\
 II. Using the $\delta$-pulse technique, we get $X\otimes I\big(H_{d}(\omega)\big)X\otimes I = \omega (X\otimes X + I \otimes X ) - Z \otimes I - Y \otimes Y - Z \otimes Z$. \\~\\
 III. By trotterizing the above with $H_{d}(\omega)$ we simulate $2\omega (X\otimes X + I \otimes X)$.\\~\\
 The Hamiltonian $2\omega (X\otimes X + I \otimes X)$ depends on the unknown parameter $\omega$ which makes the determination of its recurrence time unclear. Furthermore, the strong pulse technique $X\otimes I\big(H_{d}(\omega)\big)X\otimes I$ does not generate $-H_{d}(\omega)$ since $\omega (X\otimes X + I \otimes X)$ commutes with $X\otimes I$.  Therefore, we can not control with the commutator formula.\\
 It suggests that a similar procedure of polynomial approximation can not be followed for the system B. Hence, the robust controllability of system B remains undecided.
 
 \section{A numerical approach for robust controllability} \label{NA}
 
We now consider a numerical approach to determine robust controllability. We discretize the unknown parameter $\omega$ within an interval $[\omega_{0}, \omega_{1}]$ and seek an optimal pulse that works for each discrete value of $\omega$.
 
Specifically, let us consider the finite set $\Omega^{N} = [\omega^{(1)}, \omega^{(2)}...\omega^{(N)}] \subset [\omega_{0}, \omega_{1}]$; Our aim is to use a suitable pulse $f(t)$ to send each of the individual system $H_{(\omega^{n})}(t) = H_{d}(\omega^{n}) + f(t)H_{c}$, where $n = 1, 2, 3, ...N$ to a target unitary $U_{targ}$ at the same time t. \\
The ensemble can be represented by a single larger dimensional system with $\bar{H}_{d}(\Omega^{N}) = \bigoplus\limits_{n=1}^{N} H_{d}(\omega^{n})$ and $\bar{H}_{c} = \bigoplus \limits_{n=1}^{N} H_{c}$ in a block-diagonal form. 
\begin{equation} \label{eq0}
\bar{H}_d(\omega^{N}) = 
\begin{pNiceArray}{>{\strut}cccc}[margin,extra-margin = 3pt]
H_{d}(\omega^{1}) &  & \\
 & H_{d}(\omega^{2}) & \\
 &  \ddots & &\\
 & & &  H_{d}(\omega^{N}) \\
\end{pNiceArray}.
\end{equation}
The controllability of this larger system can be determined using the LARC condition. It will be characterized by the maximal number of linearly independent Hamiltonians generated by the multiple commutators of $\bar{H}_{d}(\Omega^{N})$ and $\bar{H}_{c}$. The maximum number of these linearly independent Hamiltonians is $N(d^{2}-1)$, where $d$ is the dimensional size of the individual system.

The robust controllability for such an ensemble is guaranteed if \cite{gabrielJPAMGoptimal,dirrMCSSaccess,mohamedEJCensemble},\\~\\
1. Each system $H_{(\omega^{n})}(t)$ in the system is individually fully controllable. Full controllability is characterized by the LARC criterion. \\~\\
2. All Hamiltonians $H_{(\omega^{k})}(t)$ and $H_{(\omega^{j})}(t)$ are not mutually Lie-related.

Given two pairs $(A, B) \in \mathfrak{g}$ and $(A', B') \in \mathfrak{g'}$ for any two arbitrary Lie-algebras $\mathfrak{g}$ and $\mathfrak{g'}$, if there exists a Lie-algebra isomorphism $\phi:\mathfrak{g}\rightarrow\mathfrak{g'}$  such that $\phi(A) = A'$ and $\phi(B) = B'$, then two pairs $(A, B)$ and $(A', B')$ are Lie-related.

While the first condition guarantees the implementation of an arbitrary quantum gate on the individual systems, the second condition implies independent implementation of quantum gates on each of the individual systems. The second condition leads to the ensemble controllability, where we can implement $\bar{U} = \bigoplus \limits_{n=1}^{N}U_{n}$ with $U_{n}$ being any unitary in $SU(d)$. A special case of ensemble control where $U_{n} = U$ $\forall n$, implies robust control. 

While in principle, we can increase the number of discretizations arbitrarily in the numerics, this also expands the size of the ensemble we aim to manipulate. The increment in the ensemble size could,  in turn, arbitrarily increase the minimum time required to control the entire ensemble. Hence, it is difficult to determine whether this scheme works in the continuous limit. The scaling of the minimum control time with the ensemble size emerges as a crucial aspect in deciding the controllability of the system.  To investigate this scaling using numerical optimization, we estimate the minimum control time $T_{\varepsilon}$ for $N$ discretization within an allowed target error $\varepsilon$. Duhamel's principle (see, e.g., \cite{ArenzNJPtheroles}),

\begin{equation} \label{ineq1}
\Vert U_{\omega}(t) - U_{\sigma}(t) \Vert \leq t ||H_{d (\omega)} - H_{d (\sigma)}||,
\end{equation}
where $U_{\omega}(t) = \hat{T}e^{-i\int \limits_{0}^{t}H_{\omega}(\tau) d\tau}$, suggests that if the scaling of $T_{\varepsilon}$ is less than $\mathcal{O}(N)$, the worst error in between the $N$ points becomes smaller as $N$ increases \cite{ryosukePRArobust}. Hence, robust controllability for a continuous unknown parameter can be demonstrated by checking the scaling $T_{\varepsilon}$ with $N$. We use gradient ascent pulse engineering(GRAPE)\cite{KHANEJA2005296, benjamin2012}, a gradient-based optimization algorithm for our numerical approach, utilizing the QuTip package\cite{JOHANSSON20131234, JOHANSSON20121760} for our numerical simulations. However, a potential problem is that with growing matrix size, the optimization becomes inefficient.
 
\begin{figure}[h!]
\begin{center}
\begin{minipage}{\linewidth}
\includegraphics[width=\textwidth]{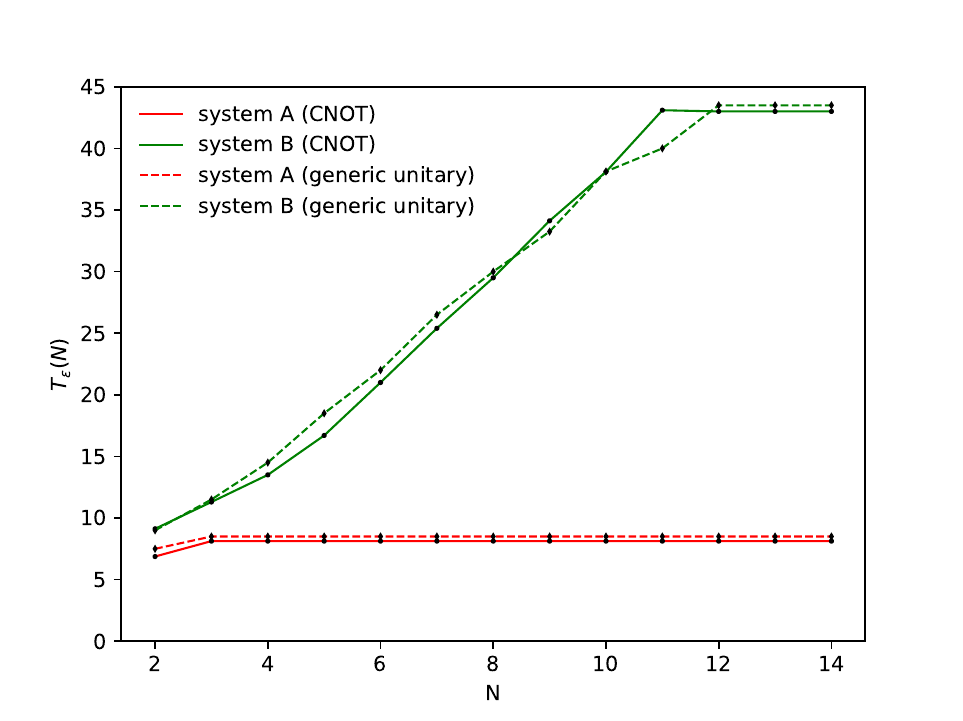}
\caption{The figure shows how the minimum control time $T_{\varepsilon}(N)$ scales with the ensemble size $N$ for System A and System B. The black dots on the curve depict the minimum control time $T_{\varepsilon}(N)$ for a target error $\varepsilon = 10^{-3}$ within $\omega \in [1, 2]$. Solid lines indicate the scaling behaviour for the CNOT gate and dashed lines correspond to the scaling for the generic unitary $U$ given in Eq. [\ref{generic}].}
\label{scaling}
\end{minipage}
\end{center}
\end{figure}

\subsection{Optimization process}\label{NR}

The numerical scheme aims to find an optimized pulse $ f(t)$ that effectively implements a target unitary for each system in the ensemble. We utilize the quantum control library of QuTip for numerical calculations. The process involves defining several parameters within the module: control time (total pulse duration), the number of segments to divide the pulse into segments of equal length, a drift Hamiltonian $ H_d $, a control Hamiltonian $ H_c $, a target unitary, and a target infidelity value which serves as the figure of merit for the algorithm. 

The quality of the pulse $f(t)$ and the controllability are assessed by calculating the infidelity $ \varepsilon(\omega) $ between the evolved unitary $U_\omega(T)$ generated by $ H_d(\omega) + f(t)H_c $ after an evolution time $ T$ and the target unitary. \\
The ensemble is chosen as a finite set of equally spaced discretized values within a closed interval $ \Omega_N = [\omega^{(1)}, \omega^{(2)}, \dots, \omega^{(N)}] \subset [\omega_0, \omega_1] $. For all numerical simulations in this work, the interval is chosen to be $ [\omega_0, \omega_1] = [1, 2] $. 

The numerical simulation optimizes for a pulse $ f(t) $ to achieve a target infidelity value. We use a gradient ascent method, GRAPE, to find the optimized pulse. In this procedure, we start with a piecewise constant initial random control pulse. The pulse parameters, such as the amplitude, are then varied iteratively until a desired minimum in the control manifold is achieved. The descent direction on the control manifold is determined by the gradient of the infidelity with respect to the pulse amplitude. The gradient is calculated using the limited-memory Broyden-Fletcher-Goldfarb-Shanno (L-BFGS) method \cite{Fouquieres2011, PRXQuantum, Liu1989, numerical2006, broyden1970, fletcher1970, Shanno1970, Goldfarb1970}. This method estimates the Hessian matrix of the second derivative of the infidelity and uses it to approximate the control manifold as a parabola, thereby finding the local minima of the infidelity. GRAPE then provides an optimized piecewise constant pulse that yields the best fidelity for a given pulse duration. Different parameters such as pulse duration, number of iterations, number of segments for the pulse, and type of initial guess pulse can be varied within the GRAPE through the QuTip module to achieve the desired output.

In our optimization process, to calculate the fidelity between the target unitary and the optimized unitary, we employ an average phase-sensitive fidelity. For closed quantum system, the phase-sensitive fidelity is defined as \cite{MachnesPRAcomparing, PhysRevA.72.042331},
\begin{equation}\label{su}
f_{SU} = \frac{1}{N}Re\hspace{0.1cm}tr\{U_{targ}^{\dagger}U(T)\},
\end{equation}
where SU stands for the special unitary group SU$(N)$.

\subsection{Results and robust controllability}

\begin{figure}[h!] 
\begin{center}
\begin{minipage}{\linewidth}
\includegraphics[width=\textwidth]{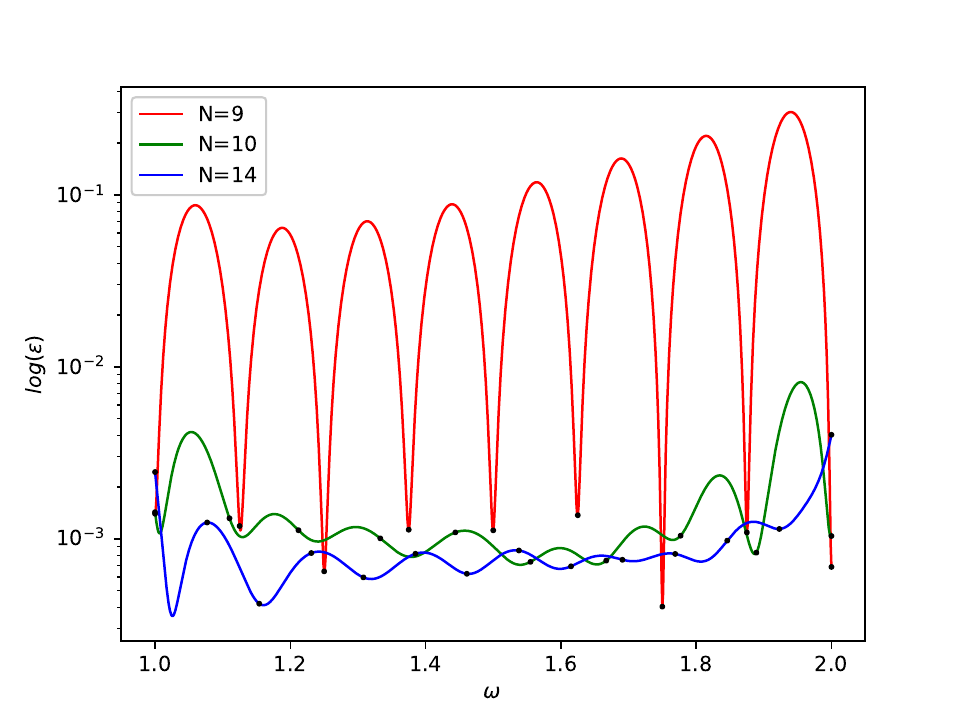}
\caption{The figure shows how an optimized pulse for system B behaves in between omega for three different values of discretization $N = 9, 10$ and $14$, with a target CNOT. We can see that for $N>9$, the error for all $\omega \in [1, 2]$ is suppressed significantly and for $N=14$ we get an optimized pulse for which the error lies below $10^{-3}$. The black dots indicate errors for the points in $\Omega_{N}$. }
\label{betweenomega}
\end{minipage}
\end{center}
\end{figure}

We investigate the robust controllability of the system B using this numerical approach. System B exhibits the two key properties: full controllability for any particular $\omega$, and for $\omega \neq \omega_0$, $H_\omega$ and $H_{\omega_0}$ are not mutually Lie-related, which suggests that for any finite discretization we can search for a pulse $f(t)$ that provides robust controllability for this system within an arbitrary error. With these conditions in place, we search for a robust control pulse for an error of $\varepsilon = 10^{-3}$ using numerical optimization.\\
In the numerical approach, we first examine the scaling of the minimum control time $T_\varepsilon$ with the number of discretizations $N$. The scaling provides crucial insight into robust controllability, as evident by Eq.~[\ref{ineq1}]. Fig.~\ref{scaling} plots the scaling for both CNOT and a random unitary as the target and compares the scaling behaviour of system A~\cite{ryosukePRArobust} with system B, which reflects their controllability. For a comprehensive numerical assessment of controllability, it is sufficient to use CNOT and a generic two-qubit unitary as target operations for two-qubit systems. We consider the following generic unitary matrix, generated from a single sample drawn from the Haar measure on the unitary group, which remains fixed throughout the numerical analysis:
\begin{widetext}
\begin{align} \label{generic}
U=
\begin{pmatrix}
0.56608 + 0.00933i & 0.09906 + 0.05347i & -0.02898 - 0.30192i & 0.20154 + 0.73087i \\
0.17824 + 0.01578i & 0.88373 - 0.03802i & 0.17237 + 0.38526i & -0.02653 - 0.08195i \\
0.57611 - 0.27732i & -0.41590 + 0.07183i & 0.20749 + 0.58112i & -0.14760 - 0.10256i \\
0.24404 + 0.42318i & -0.06879 + 0.14844i & 0.53465 - 0.25151i & 0.46820 - 0.40776i
\end{pmatrix}
\end{align}
\end{widetext}
The scaling for system A, demonstrably robustly controllable through theoretical analysis, plateaus very quickly, reaching this point with a discretization of $N = 3$. In contrast, system B requires a longer time scale and more discretization points $N \geq 10$ to achieve a similar plateau. The plateauing behaviour, as suggested by Eq.~[\ref{ineq1}], indicates rboust controllability of the system B. We observed a similar scaling for both CNOT and the generic unitary $U$.  With this observation, for the remainder of this paper, we restrict our numerical analysis to CNOT alone due to demanding computational effort.  \\
We further evaluate robust controllability by analyzing the performance of an optimized pulse between the optimized points in $[\omega_{0}, \omega_{1}]$. Specifically, whether we can obtain an optimized pulse that maintains error under $\varepsilon = 10^{-3}$ within $[\omega_{0}, \omega_{1}]$. Fig.[\ref{betweenomega}] displays that the worst error between the optimized decreases as we increase the number of discretization $N$. For $N>9$, the error in between the optimized points is significantly supressed and for $N=14$, the error remains mostly below $10^{-3}$.   

 Following the numerical results for the scaling of control time with $N$ and searching for a pulse that significantly reduces errors for all points in $[\omega_{0}, \omega_{1}] = [1,2]$ close to $\varepsilon = 10^{-3}$, suggests that the system B is indeed robustly controllable. However, numerical results do not guarantee existence of an optimized pulse that can achieve arbitrarily low error. It is worth noting that the numerical search for optimized pulses and the controllability for system B is, in general, way more difficult compared to system A \cite{ryosukePRArobust}. This leads us to further explore the controllability behaviour in the two systems.
 
 \begin{figure*}
\centering
\hspace*{\fill}
\includegraphics[width=0.486\textwidth]{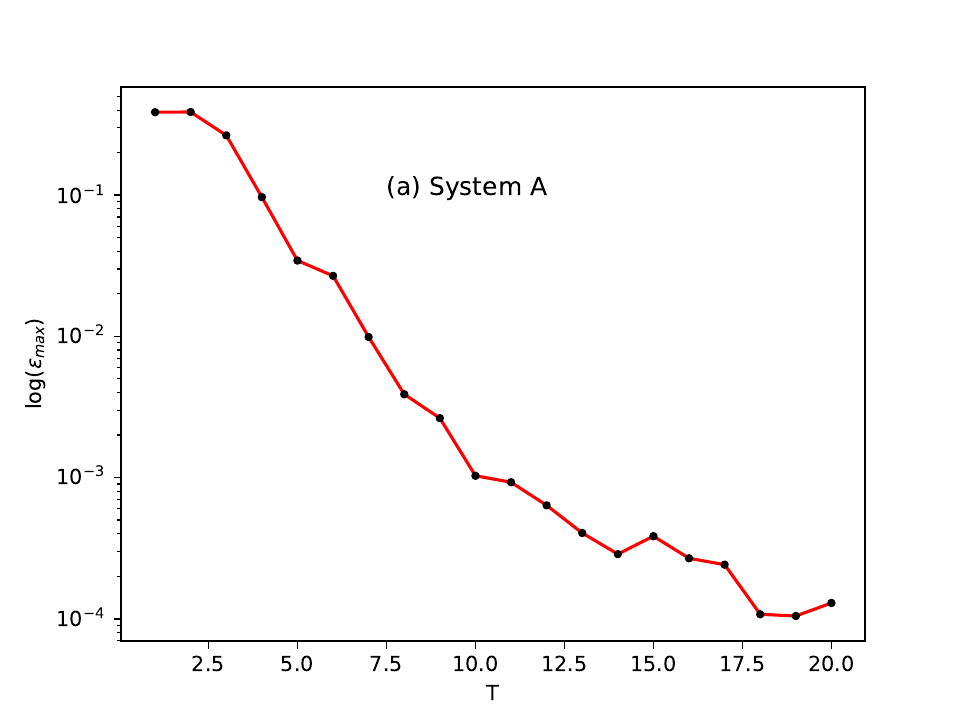} 
\hspace*{\fill} 
\includegraphics[width=0.446\textwidth]{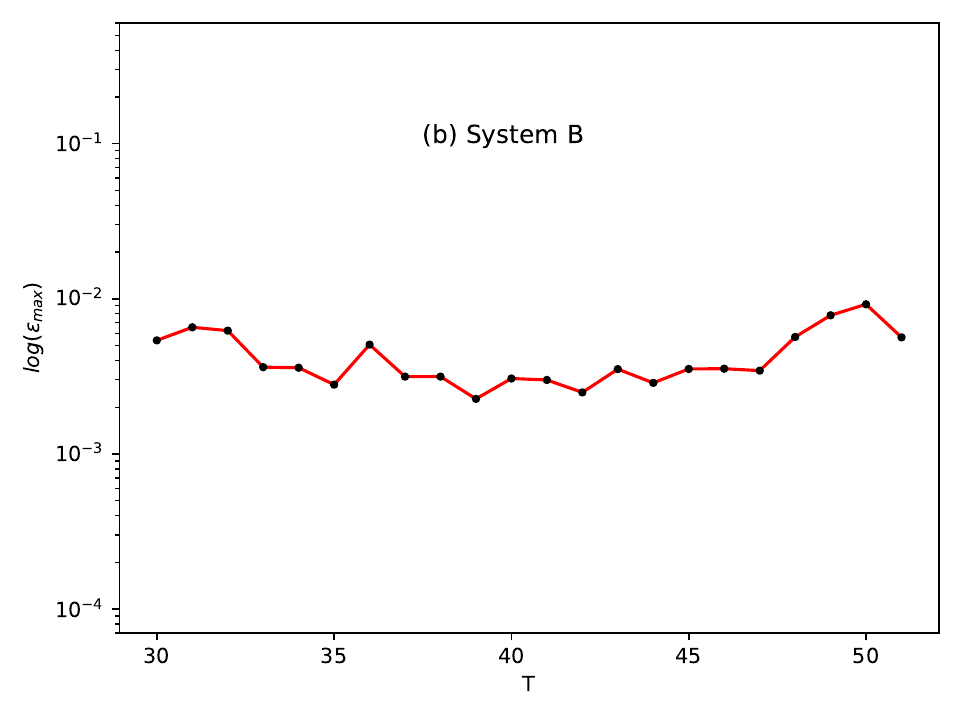} 
\hspace*{\fill}
\caption{(a) System A. (b) System B. The above figure illustrates how worst-case error behaves for system A and system B as a function of control time $T$, with the number of discretizations set to $N=12$ for both cases. For system A, we observe a monotonic decrease in worst error as the control time increases, in contrast, system B exhibits an approximatley constant behaviour with increasing control time.}
\label{fragility}
\end{figure*}

\subsection{Comparison of controllability with varying control time}\label{PA}
We observe from the numerical results that for system B, the error scaling with the minimum control time $T_{min}$ plateaus more slowly compared to system A.  This suggests a fundamental difference in their controllability. To further explore this, we examine how both systems behave as the control time increases.

To do so, we analyze the worst-case error for a given optimized pulse within the range $[\omega_0, \omega_1]$ for a chosen control time $T$ and visualize this as control time varies for both systems, comparing their behaviour. Instead of increasing both $N$ and $T$ (which, as shown in Fig. [\ref{scaling}], demonstrates that both systems are robustly controllable), we keep N constant and increase T. Surprisingly, system A improves, whereas system B remains roughly the same. 

We choose a discretization $N =12$ and CNOT as the target unitary and we plot the results as shown in Fig.[\ref{fragility}]. The choice of $N=12$ is motivated by the fact that the scaling of the $T_{min}$ with N already saturates for both the systems, as shown in Fig [\ref{scaling}]. For system A, the minimum control time required to achieve an average error of $10^{-3}$ for $N=12$ is $T_{min}=8$, while for system B, it is $T_{min}= 40.1$. We plotted the worst error as a function of control time T, taking values both below and above $T_{min}$. 

Fig.[\ref{fragility}] illustrates that system A exhibits a well-behaved trend, with the worst error monotonically decreasing as $T$ increases. In contrast, system B behaves differently as the worst errors remain nearly constant with increasing control time, showing no significant improvement. This highlights the observation that, while both systems exhibit robust controllability based on the numerical results, their controllability properties differ significantly. Notably, system A is much easier to control. 

In the next section, we further enhance our numerical approach by incorporating a penalty term in the fidelity function, allowing for deeper comparison of the controllability of the two systems.

\section{Improvement of the robustness of the pulses by modification of the fidelity}\label{MF}
In the numerical technique, we discretize the unknown parameter into a finite set and optimize for those specific set points. The fidelity changes sharply from the non-optimized points towards the optimized points, causing $|\frac{\partial f}{\partial \omega}|$ to acquire large values at these points as shown in Fig.[\ref{betweenomega}]. Consequently, the pulses generate high errors between the optimized points, leading to overall substandard performance in terms of robustness. We aim to improve the robust pulses by emplying a modified fidelity that will provide better fidelity between the optimized points, keeping the errors well below the target $\varepsilon$ for all points of $\omega \in [\omega_{0}, \omega_{1}]$.

\subsection*{New fidelity}
To enhance the quality of the robust pulse, we define a new fidelity function that can provide better fidelity in between the non-optimized points. We achieve this by introducing a penalty term into the fidelity as,
\begin{equation}\label{newfid}
f' = f - \alpha |\frac{\partial f}{\partial \omega}|.
\end{equation}
The second term on the right, $ |\frac{\partial f}{\partial \omega}| $, acts as a penalty term and $\alpha$ serves as a positive weight on the penalty. With the introduction of this modified fidelity, the optimizer tries to reduce $|\frac{\partial f}{\partial \omega}|$ while searching for an optimized pulse. This helps to prevent rapid changes in fidelity from non-optimized points towards the optimized points. 

{\it Estimation of $|\frac{\partial f}{\partial \omega}|$: } To estimate the new fidelity in our optimization, we first derive an analytical expression for $\frac{\partial f}{\partial \omega}$.

The control problem we are solving is bi-linear control problem and can be expressed as a Schr\"{o}dinger equation of a unitary evolution operator,
\begin{equation}\label{bilinear}
\dot{U}(t) = -i  \Big( H_d + \sum\limits_{j=1}^{m}u_{j}(t)H_{j} \Big)U(t).
\end{equation}
Where $H_d$ is the drift Hamiltonian and $u_{j}$'s are the control amplitudes corresponding to the control Hamiltonians $H_j$. Let us define a Hamiltonian by considering an unknown parameter within the drift Hamiltonian
\begin{equation}\label{hamiltonian}
H_{u}(t, \omega) = H_d (\omega) + \sum_{j = 1}^{m} u_{j}(t) H_{j} =( \omega H_{1} + H_{2} ) + \sum_{j = 1}^{m} u_{j}(t) H_{j}.
\end{equation}
Here, we divide the drift Hamiltonian into two components: $H_{1}$, which is solely dependent on the unknown parameter $\omega$ and $H_{2}$. 
For an evolution involving a piecewise constant pulse with equal time interval $\Delta t$, the unitary evolution operator for a single time slice $\Delta t$ is given by:
\begin{equation}
X = \exp\{-i \Delta t H_{u}\} = \exp\Big\{-i \Delta t(H_d + \sum_{j}u_{j}H_{j})\Big\} 
\end{equation}
Let us write the following equation to calculate the gradient of a unitary with respect to $\omega$ for a piecewise control:
\begin{eqnarray}
\frac{\partial X}{\partial \omega} && = \frac{\partial }{\partial k} \hspace{0.1cm} \exp\Big\{-i \Delta t \Big( (\omega+k) H_{1} + H_{2}  +\sum_{ j} u_{j}H_{j}\Big) \Big\} \Big |_{k=0} \nonumber \\ && = \frac{\partial}{\partial k} \hspace{0.1cm} \exp \Big\{ -i \Delta t (H_{u}(\omega) + k H_{1})\Big\}\Big |_{k=0}.
\end{eqnarray}
This expression can be estimated using the spectral theorem and by calculating the matrix functions via the eigen-decomposition. The final expression can be written as,
\begin{equation}
\frac{\partial X(t_{k}, \omega)}{\partial \omega} =  \langle \lambda_{l} |B |\lambda_{m}\rangle \hspace{0.1cm} \frac{\Big(e^{\lambda_{l}} - e^{\lambda_{m}}\Big)}{\lambda_{l}-\lambda_{m}}
\end{equation}
for $\lambda_{l} \neq \lambda_{m}$ and 
\begin{equation}
\frac{\partial X(t_{k}, \omega)}{\partial \omega} =  \langle \lambda_{l} |B |\lambda_{m}\rangle e^{\lambda_{l}} 
\end{equation}
when $\lambda_{l} = \lambda_{m}$.
Here, $| \lambda_{l}\rangle$ represents the eigenbasis of  $A = -i \Delta t \hspace{0.01cm} H_{u} $ and $B = -i \Delta t \hspace{0.01cm} H_{1}$ with $\lambda_{l}$'s denoting the corresponding eigenvalues.\\
Now, finally the expression for the gradient of the fidelity is given by,
\begin{eqnarray}\label{fidgrad}
 \frac{\partial f_{PSU}(X_{k}, \omega)}{\partial \omega} &&  =  \\ &&  \sum_{k=1}^{M}\frac{1}{N}Re \hspace{0.1cm} tr \Big[e^{-i\phi_{g}}\Lambda_{M+1:k+1}^{\dagger} \Big(\frac{\partial X_{k}}{\partial \omega}\Big)X_{k-1:0}\Big].\nonumber
\end{eqnarray}
Where $ \Lambda^{\dagger}_{M+1:k+1} = U_{target}^{\dagger}X_{M}X_{M-1}...X_{k+1}$ and $ X_{k:0} = X_{k}X_{k-1}...X_{0}$. The PSU fidelity which is a phase-insensitive is defined as~\cite{MachnesPRAcomparing, PhysRevA.72.042331}, 
\begin{align}
f_{\text{PSU}} = \frac{1}{N} \big| \text{tr} \{ U^\dagger_{\text{tar}} U(T) \} \big|, 
\end{align}
where PSU stands for projective unitary grouo PSU$(N)$. \\~\\
$X_{k}$ governs the time evolution withinin the time slice $(t_{k-1}, t_{k}]$ and is defined as:
\begin{equation}
X_{k} = e^{-i\Delta t H_{u}} = \exp\Big\{-i\Delta t(H_d + \sum_{j}u_{j}(t_{k})H_{j})\Big\}. 
\end{equation}
Here, $\Delta t = t_{k} - t_{k-1}$ for all time slices $k = 1, 2...M$ and $T = M \Delta t$. The boundary conditions are specified as $X(0) = X_{0}$ and $X_{M+1} = X_{target}$. The evolution is given by discretized evolution. A detailed analytical derivartion of $ \frac{\partial f_{PSU}(X_{k}, \omega)}{\partial \omega}$ can be found in the Appendix~\ref{A}.\\
\subsection*{Modification of pulses}
To observe the improvements achieved in pulse with the newly defined fidelity, we first obtain a robust pulse with a certain discretization $N$ within the range $[\omega_{0}, \omega_{1}]$, a control time $T$ and a target error $\varepsilon$. The pulse is obtained by setting $\alpha = 0$ in the optimization. It then serves as an initial guess pulse for a re-optimization process by setting a non-zero value for $\alpha$. While we utilize the analytical expression to compute $|\frac{\partial f}{\partial \omega}|$, a numerical approximation is used for $\frac{\partial f}{\partial u_{j}}$ to update the pulse amplitudes within the GRAPE optimizer. We evaluate the improvements for both systems A and B, which also offers insight into the inherent differences in controllability between the two systems.

\begin{figure}
\begin{center}
\begin{minipage}{\linewidth}
\includegraphics[width=\textwidth]{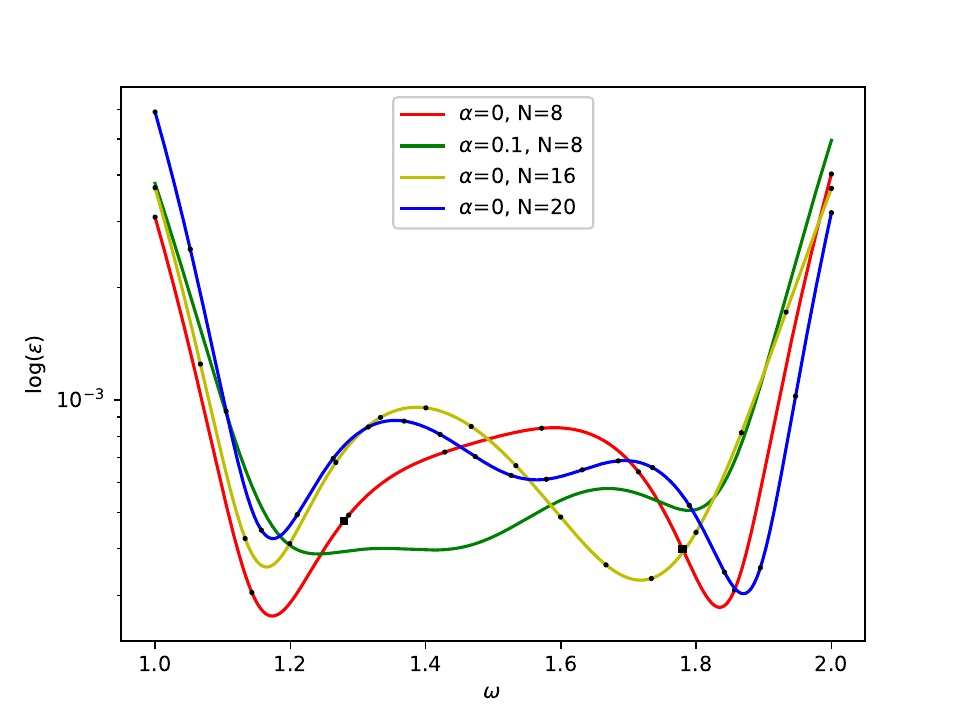}
\caption{The figure above compares errors produced by the modified pulses using the new fidelity function for system A. The red curve illustrates errors between optimized points for the optimized pulse with $N = 8$, $T = 15$, and $\alpha = 0$. The green curve represents the pulse obtained by re-optimizing the red pulse using the new fidelity with $\alpha = 0.1$ with two chosen points to reduce the gradient $|\frac{\partial f}{\partial \omega}|$. These two points are shown by the black squares on the red curve. The yellow and blue curves show errors for optimized pulses with discretizations $N = 16$ and $N = 20$, respectively, for $\alpha = 0$. We set the target error at $\varepsilon = 10^{-3}$ for the optimizations. The small black dots on each curve indicate errors at the optimized points.}
\label{hamAmod}
\end{minipage}
\begin{minipage}{\linewidth}
\includegraphics[width=\textwidth]{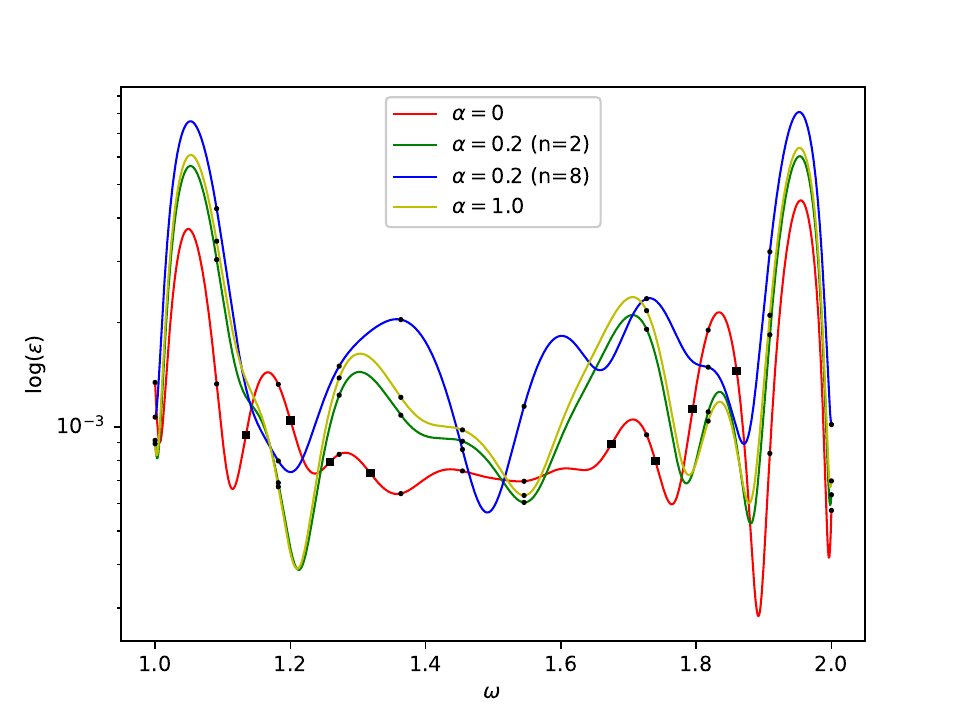}
\caption{The figure above compares errors produced by the modified pulses using the new fidelity function for system B. The red curve illustrates errors between optimized points for the optimized pulse with $N = 12$, $T = 45$, and $\alpha = 0$. The green curve depicts errors for the pulse obtained through reoptimization using the new fidelity with $\alpha = 0.2$, focusing on two points to reduce the gradient $|\frac{\partial f}{\partial \omega}|$. The blue curve represents a pulse obtained by re-optimizing with $\alpha = 0.2$, targeting eight points to reduce the gradient. The yellow curve shows a re-optimized pulse with $\alpha = 1.0$ with two points chosen to reduce the gradient. We set the target error at $\varepsilon = 10^{-3}$ for the optimizations. The small black dot on each curve indicates errors at the optimized points, while the black squares denote eight points where we aim to reduce $|\frac{\partial f}{\partial \omega}|$. When only two points are chosen, they correspond to the leftmost black and rightmost black squares.}
\label{newhamBmod}
\end{minipage}
\end{center}
\end{figure}

For system A, we first obtain an optimized pulse with $N=8, T=15$, $\varepsilon=10^{-3}$ and $\alpha = 0$. We observe that the re-optimized pulse with $\alpha = 0.1$ produces better errors between optimized points within $[\omega_{0}, \omega_{1}] = [1,2]$. This improvement is evident in Fig.[\ref{hamAmod}], where the red curve depicts errors for the optimized pulse with $\alpha=0$, while the green curve represents the re-optimized pulse with $\alpha=0.1$. We choose $n=2$ to reduce the gradient $\big|\frac{\partial f}{\partial \omega}\big|$, where $n$ indicates the number of points where we aim to reduce the gradient. These points are shown by black squares on the red curve. 

We further compare the improvements of this pulse with other pulses obtained by increasing the number of discretization while keeping the control time same. The behaviour of these pulses is illustrated by the yellow curve$(N=16)$ and blue curve$(N=20)$ in Fig.[\ref{hamAmod}]. Notably, the pulse employing the new fidelity provides better errors in between optimized points, even when compared to the errors produced by the pulses with increased discretization. This indicates that this method is superior in generating robust pulses compared to the approach that increases discretization.

For system B, we followed a similar procedure. First, we obtained an optimized pulse with $N=12, T=45,$ $\varepsilon=10^{-3}$ and $\alpha = 0$, shown by the red curve in in Fig.[\ref{newhamBmod}]. We re-optimized this pulse to enhance the behaviour in between the optimized points within $[\omega_{0}, \omega_{1}] = [1,2]$. We first chose $n=2$ to reduce the gradient, $\big|\frac{\partial f}{\partial \omega}\big|$ and re-optimized for two different values of $\alpha = 0.2, 1.0$. As we can see, this resulted in poorer performance in between optimized points, ilustrated by the green ($\alpha=0.2$) and yellow ($\alpha=1.0$) curves. To further investigate, we increased the number of points to $n=8$ to reduce the gradient while keeping $\alpha=0.2$ the same. However, as displayed in the blue curve, this failed to yield any improvements in the pulse's performance. 

Following the trend above, the observed difficulty in improving the robustness of system B's optimized pulse suggests that achieving arbitrary low error across all points within $[\omega_0, \omega_1]$ might be challenging, if not impossible. Overall, inherent differences between system A and system B is clearly evident. This is a clear indication that control of system B is in general a lot challenging than system A, a fact already observed in the previous section.

\section{Summary and conclusion} \label{SD}
We considered two systems A and B with an unknown parameter, a common scenario in real quantum systems, and compared the challenges related to their practical implementability by investigating their robust controllability through both theoretical and numerical approaches.  Within a theoretical framework, using Lie algebraic approach, system A is shown to be robustly controllable for a continuous and compact subset of the unknown parameter, however, the same approach was not able to conclusively demonstrate robust controllability for system B, essentially because we could not invert the drift.

Next, we employed a numerical approach to assess robust controllability by discretizing the unknown parameter $\omega$ within a finite set $\Omega_N$ and optimizing a pulse that  effectively works across individual systems. We examined the scaling of minimum control time $T_{\text{min}}$ with the discretization $N$ to predict robust controllability. Additionally, we modified the fidelity function with a penalty term to improve pulse robustness by reducing errors between optimized points.

Employing the numerical approach, we found a control pulse for system B that maintains errors below $10^{-3}$ for $\omega \in [1, 2]$. However, while modifying the fidelity function enhanced pulse robustness for system A, it failed to further improve system B's performance. This raises an intriguing curiosity, although the numerical approach confirms that both systems are robustly controllable, their optimization behaviour differs qualitatively. In particular, compared to system A, system B exhibits a slower plateau, no improvement in pulse performance between optimized points with the modified fidelity, and worst-case errors remaining roughly unchanged as the control time increases. This emphasizes the greater difficulty in controlling system B. 

A deeper mathematical analysis is required to fully understand these differences between the two systems beyond just the question of robust controllability. Notably, one key factor contributing to these differences in the numerical results may be the sign inversion of the drift Hamiltonian.

The protocol developed here for assessing robust controllability and optimizing error-resilient pulses holds potential for high-fidelity gate implementation in practical quantum systems, including superconducting qubits used by IBM and Google, or trapped-ion qubits.  

\section*{Acknowledgements}
A.D. is supported by the primary PhD scholarship from Sydney quantum academy (SQA).

\newpage
\bibliographystyle{unsrt}
\bibliography{ref}

\appendix
\section{Calculation of the gradient of fidelity with respect to $\omega$}\label{A}
We want to define a modified fidelity which is written as,
\begin{equation}
f(t, \omega) = f(t) - \alpha \mid \frac{\partial f(t,\omega)}{\partial \omega} \mid.
\end{equation}

Let us first define a quantity

\begin{equation}
g = \frac{1}{N} tr\{ U^{\dagger}_{target} U(T) \}. 
\end{equation}

We next define,
\begin{equation}
X_{k} = e^{-i\Delta t H_{u}} = \exp\Big\{-i\Delta t(H_d + \sum_{j}u_{j}(t_{k})H_{j})\Big\}, 
\end{equation}
where $X_{k}$ governs the controlled time evolution in the time slice $(t_{k-1}, t_{k}]$, $\Delta t = t_{k} - t_{k-1}$ for all time slices $k = 1, 2...M$ and $T = M \Delta t$. The boundary condition is given by,  $X(0) = X_{0}$ and $X_{M+1} = X_{target}$. 
We discretize the evolution and write the following equation;
\begin{equation}
X(t_{k}) = X_{k:0} = X_{k}X_{k-1}...X_{0}.
\end{equation}

Now for the phase independent fidelity with $\hat{X} = \bar{X} \otimes X $,

\begin{eqnarray}
f^{2}_{PSU} && = \frac{1}{N^{2}}Re \hspace{0.1cm} tr\{\hat{U}_{target}^{\dagger}\hat{X}(T)\}   \nonumber \\ && = \frac{1}{N^{2}}Re \hspace{0.1cm} tr \{(U^{t}_{target}\bar{X}_{T})\otimes (U_{target}^{\dagger}X_{T})\} \nonumber \\ && =  \frac{1}{N^{2}} |tr\{U^{\dagger}_{target}X_{T}\} |^{2} = |g|^{2}
\end{eqnarray} 
From above equation it follows,
\begin{equation}
f_{PSU} = \frac{1}{N} |tr\{U^{\dagger}_{target}X_{T}\} | = \frac{1}{N}|tr \{\Lambda^{\dagger}_{M+1:k+1}X_{k:0}\}|,
\end{equation}
where $ \Lambda^{\dagger}_{M+1:k+1} = U_{target}^{\dagger}X_{M}X_{M-1}...X_{k+1} $.\\
From $f^{2}_{PSU}(\omega) = |g(\omega)|^{2}$ we can write,
\begin{eqnarray}
&& \frac{\partial}{\partial \omega} f^{2}_{PSU}(\omega) = \frac{\partial}{\partial \omega} |g(\omega)|^{2} \nonumber \\  \Longrightarrow  && 2 f_{PSU}(\omega) \frac{\partial}{\partial \omega} f_{PSU}(\omega) \nonumber \\ && = 2 |g(\omega)| \frac{\partial |g(\omega)|}{\partial \omega} = 2 |g(\omega)| \frac{\partial}{\partial \omega} f_{PSU}(\omega) \nonumber \\ \Longrightarrow && \frac{\partial}{\partial \omega} f_{PSU}(\omega) = \frac{1}{2|g(\omega)|}\frac{\partial}{\partial \omega}f^{2}_{PSU}(\omega)\nonumber \\ &&
\end{eqnarray}

This leads us to find $\frac{\partial}{\partial \omega}f^{2}_{PSU}(\omega)$. The quantity $f^{2}_{PSU}$ consisted of  $X(t_{k}, \omega)$ 
\begin{eqnarray}
&& \frac{\partial f^{2}_{PSU}(X_{t_{k},}, \omega)}{\partial \omega}  \nonumber \\ && =  \frac{1}{N^{2}} \frac{\partial}{\partial \omega} Re \hspace{0.1cm} tr \{\Lambda_{M+1:k+1}^{t}\bar{X}_{k:0}\otimes \Lambda_{M+1:k+1}^{\dagger}X_{k:0} \} \nonumber \\  &&
\end{eqnarray}

Let us first do a derivation for the above derivative by taking $X_{k}$ as a function of some parameter $u_{k}$ with $ \Lambda^{\dagger}_{M+1:k+1} = U_{target}^{\dagger}X_{M}X_{M-1}...X_{k+1}$ and $ X_{k:0} = X_{k}X_{k-1}...X_{0}$.
\begin{eqnarray}
&& \frac{\partial f^{2}_{PSU}(X_{t_{k}}, u_{t_{k}})}{\partial u_{t_{k}}}  \nonumber \\ && = \frac{1}{N^{2}} \frac{\partial}{\partial u_{t_{k}}} Re \hspace{0.1cm} tr \{\Lambda_{M+1:k+1}^{t}\bar{X}_{k:0}\otimes \Lambda_{M+1:k+1}^{\dagger}X_{k:0}, \} \nonumber \\ && = \frac{1}{N^{2}} Re \hspace{0.1cm} tr \{\Lambda_{M+1:k+1}^{t}\Big(\frac{\partial \bar{X}_{k}}{\partial u_{t_{k}}}\Big)\bar{X}_{k-1:0}\otimes \Lambda_{M+1:k+1}^{\dagger}X_{k:0}  \nonumber \\ && + \Lambda_{M+1:k+1}^{t}\bar{X}_{k:0}\otimes \Lambda_{M+1:k+1}^{\dagger}\Big(\frac{\partial X_{k}}{\partial u_{t_{k}}}\Big)X_{k-1:0} \}, \nonumber \\ && = \frac{1}{N^{2}} Re \{tr [ \Lambda_{M+1:k+1}^{t}\Big(\frac{\partial \bar{X}_{k}}{\partial u_{t_{k}}}\Big)\bar{X}_{k-1:0} ] \hspace{0.1cm} tr [ \Lambda_{M+1:k+1}^{\dagger}X_{k:0} ]\} \nonumber \\ &&+  \hspace{0.1cm} tr [  \Lambda_{M+1:k+1}^{t}\bar{X}_{k:0}] \hspace{0.1cm} tr[ \Lambda_{M+1:k+1}^{\dagger}\Big(\frac{\partial X_{k}}{\partial u_{t_{k}}}\Big)X_{k-1:0}]\} \hspace{0.5cm}, \nonumber \\ &&  \nonumber \\ && = \frac{2}{N} Re \hspace{0.1cm} tr \big[g^{\ast}\Lambda_{M+1:k+1}^{\dagger}\Big(\frac{\partial X_{k}}{\partial u_{t_{k}}}\Big)X_{k-1:0}\big], 
\end{eqnarray}
where we have used $ tr(A\otimes B) = tr(A) tr(B)$ and $tr(A)\hspace{0.1cm} tr(B) = tr(A \hspace{0.1cm} tr(B))$.

Now let us replace $u_{k}$ by $\omega$ and also we need to keep in mind that each of $X_{k}$'s are function of $\omega$ except $X_{0} = \mathbb{I}$ and $X_{M+1} = U_{targ}$. Therefore, applying the chain rule of derivatives, we can write:
\begin{eqnarray}
&& \frac{\partial f_{PSU}(X_{t_{k}},  \omega)}{\partial \omega} \nonumber =  \frac{1}{2|g(\omega)|}\frac{\partial}{\partial \omega}f^{2}_{PSU}(\omega) \\ \nonumber && = \frac{1}{N}Re \hspace{0.1cm} tr \Big[e^{-i\phi_{g}}\Lambda_{M+1}^{\dagger}\Big(\frac{\partial X_{M}}{\partial \omega}\Big)X_{M-1:0}\Big] +..\nonumber \\ && + \frac{1}{N}Re \hspace{0.1cm} tr \Big[e^{-i\phi_{g}}\Lambda_{M+1:k+1}^{\dagger}\Big(\frac{\partial X_{k}}{\partial \omega}\Big)X_{k-1:0}\Big] +.. \nonumber \\ && + \frac{1}{N}Re \hspace{0.1cm} tr \Big[e^{-i\phi_{g}}\Lambda_{M+1:k-(M-1)}^{\dagger}\Big(\frac{\partial X_{1}}{\partial \omega}\Big)X_{0}\Big], \nonumber \\ &&  = \sum_{k=1}^{M}\frac{1}{N}Re \hspace{0.1cm} tr \Big[e^{-i\phi_{g}}\Lambda_{M+1:k+1}^{\dagger}\Big(\frac{\partial X_{k}}{\partial \omega}\Big)X_{k-1:0}\Big]. \nonumber \\ &&
\end{eqnarray}
Where $e^{-i\phi_{g}} = \frac{g^{\ast}}{|g|}$. This is done by using the polar form $g = |g|e^{-i\phi_{g}}$ which is useful for numerical formulation.\\

Let us now look at how to define finding derivative of a unitary. First of all, let us define a Hamiltonian where some part of the drift depends on an unknown parameter $\omega$. We divide the drift Hamiltonian into two parts, $H_d(\omega) = \omega H_{1} + H_{2}$. Where only $H_{1}$ depends on $\omega$. The full Hamiltonian can be written as,
\begin{equation}
H_{u}(t, \omega) = H_d (\omega) + \sum_{j = 1}^{m} u_{j}(t) H_{j} =\Big( \omega H_{1} + H_{2} \Big) + \sum_{j = 1}^{m} u_{j}(t) H_{j}.
\end{equation}
Next, we define derivative of a unitary generated by the above Hamiltonian with respect to $\omega$ for a time slice $\Delta t$,
\begin{eqnarray}
\frac{\partial X}{\partial \omega} && = \frac{\partial }{\partial k} \hspace{0.1cm} \exp\Big\{-i \Delta t \Big( (\omega+k) H_{1} + H_{2}  +\sum_{ j} u_{j}H_{j}\Big) \Big\} \Big |_{k=0}, \nonumber \\ && = \frac{\partial}{\partial k} \hspace{0.1cm} \exp \Big\{ -i \Delta t (H_{u}(\omega) + k H_{1})\Big\}\Big |_{k=0}.
\end{eqnarray}
We can now use the spectral theorem to calculate the above quantity via the eigen-decomposition.

For a pair of arbitrary Hermitian matrices A, B with $x \in \mathbb{R}$ and taking $\{|\lambda_{v}\rangle \}$ as the orthonormal eigenbasis with the eigenvalues $\{\lambda_{v}\}$ it follows

\begin{eqnarray}
D && = \langle \lambda_{l} |\frac{\partial}{\partial x} e^{A+xB}| \lambda_{m}\rangle \Big|_{u=0}, \\ &&  = \langle \lambda_{l} | \frac{\partial}{\partial x} \sum_{n =0}^{\infty} \frac{1}{n!} (A+xB)^{n}|\lambda_{m}\rangle \Big|_{x=0}
\end{eqnarray}

Since A and B are non-commuting matrices, we should take care of the matrix multiplication while taking the derivative. We can follow the procedure as,
\begin{eqnarray}
\frac{\partial}{\partial x } (A+x B)^{2} = B(A + xB) + (A + xB)B. 
\end{eqnarray}
\begin{eqnarray}
&&\frac{\partial}{\partial x} (A+xB)^{3}  \nonumber \\ && =  \frac{\partial}{\partial x} (A+xB) (A+xB)^{2} + (A+xB)\frac{\partial}{\partial x} (A+xB)^{2}, \nonumber \\ && = B(A+xB)^{2} + (A+ xB)B(A+xB)+ (A+xB)^{2}B. \nonumber
\end{eqnarray}
\begin{eqnarray}
\frac{\partial}{\partial x } (A+x B)^{4} = && (A+xB)^{3} + (A+xB)B(A+xB)^{2} + \nonumber \\ && (A+xB)^{2}B(A+xB) + (A+xB)^{3}B. \nonumber
\end{eqnarray}
A similar procedure as above enables us to write,
\begin{eqnarray}
&& \langle \lambda_{l} | \frac{\partial}{\partial x} \sum_{n =0}^{\infty} \frac{1}{n!} (A+xB)^{n}|\lambda_{m}\rangle \Big|_{f=0} \nonumber \\ && = \langle \lambda_{l} | \sum_{n =0}^{\infty} \frac{1}{n!} \sum_{q=1}^{n}(A+xB)^{q-1} B (A+xB)^{n-q}|\lambda_{m}\rangle \Big|_{x=0}, \nonumber \\ && =  \langle \lambda_{l} |  \sum_{n =0}^{\infty} \frac{1}{n!} \sum_{q=1}^{n}(A)^{q-1} B (A)^{n-q}|\lambda_{m}\rangle, \nonumber \\ && =   \sum_{n =0}^{\infty} \frac{1}{n!} \sum_{q=1}^{n} \lambda_{l}^{q-1} \langle \lambda_{l} |B |\lambda_{m}\rangle \lambda_{m}^{n-q}, \nonumber \\ && = \langle \lambda_{l} |B |\lambda_{m}\rangle \sum_{n =0}^{\infty} \frac{1}{n!} \sum_{q=1}^{n} \lambda_{l}^{q-1} \lambda_{m}^{n-q}.
\end{eqnarray}
We can arrive to two different situations here:\\
I) when $\lambda_{l} = \lambda_{m}$ ,
\begin{eqnarray}
D && = \langle \lambda_{l} |B |\lambda_{m}\rangle \sum_{n =0}^{\infty} \frac{1}{n!} \sum_{q=1}^{n} \lambda_{l}^{n-1}, \nonumber \\ && = \langle \lambda_{l} |B |\lambda_{m}\rangle \sum_{n =0}^{\infty} \frac{1}{n!} \hspace{0.1cm} n \hspace{0.1cm} \lambda_{l}^{n-1}, \nonumber \\ && = \langle \lambda_{l} |B |\lambda_{m}\rangle e^{\lambda_{l}}.
\end{eqnarray}
II) when $\lambda_{l} \neq \lambda_{m}$,
\begin{eqnarray}
D && = \langle \lambda_{l} |B |\lambda_{m}\rangle \sum_{n =0}^{\infty} \frac{1}{n!} \sum_{q=1}^{n} \lambda_{l}^{q-1} \lambda_{m}^{n-q},\nonumber \\ && = \langle \lambda_{l} |B |\lambda_{m}\rangle \sum_{n =0}^{\infty} \frac{1}{n!} \sum_{q=1}^{n} \lambda_{l}^{q-1} \lambda_{m}^{-(q-1)} \lambda_{m}^{n-1}, \nonumber \\ && = \langle \lambda_{l} |B |\lambda_{m}\rangle \sum_{n =0}^{\infty} \frac{1}{n!} \sum_{q=1}^{n} \Big( \frac{\lambda_{l}}{\lambda_{m}} \Big)^{q-1}  \lambda_{m}^{n-1}.
\end{eqnarray}
Now $\sum_{q=1}^{n} \Big( \frac{\lambda_{l}}{\lambda_{m}} \Big)^{q-1}  \lambda_{m}^{n-1}$ this forms a geometric series,
\begin{equation}
\sum_{q=1}^{n} \Big( \frac{\lambda_{l}}{\lambda_{m}} \Big)^{q-1}   = 1 + \Big(\frac{\lambda_{l}}{\lambda_{m}} \Big) + \Big(\frac{\lambda_{l}}{\lambda_{m}} \Big)^{2} + .........\Big(\frac{\lambda_{l}}{\lambda_{m}} \Big) ^{n-1}. 
\end{equation}
Using this we get,
\begin{eqnarray}
D && = \langle \lambda_{l} |B |\lambda_{m}\rangle \sum_{n =0}^{\infty} \frac{1}{n!} \hspace{0.1cm} \lambda_{m}^{n-1} \hspace{0.1cm} \frac{\Big( \frac{\lambda_{l}}{\lambda_{m}}\Big)^{n}-1}{\Big( \frac{\lambda_{l}}{\lambda_{m}}\Big)-1}, \nonumber \\ && = \langle \lambda_{l} |B |\lambda_{m}\rangle \sum_{n =0}^{\infty} \frac{1}{n!} \hspace{0.1cm} \lambda_{m}^{n-1} \hspace{0.1cm} \frac{\Big( \lambda_{l}^{n} - \lambda_{m}^{n}\Big)}{\Big( \lambda_{l} - \lambda_{m} \Big)} \hspace{0.1cm} \frac{\lambda_{m}}{\lambda_{m}^{n}}, \nonumber  \\ && = \langle \lambda_{l} |B |\lambda_{m}\rangle \sum_{n =0}^{\infty} \frac{1}{n!} \hspace{0.1cm} \hspace{0.1cm} \frac{\lambda_{l}^{n} - \lambda_{m}^{n}}{ \lambda_{l} - \lambda_{m}}, \nonumber \\ && = \langle \lambda_{l} |B |\lambda_{m}\rangle \hspace{0.1cm} \frac{\Big(e^{\lambda_{l}} - e^{\lambda_{m}}\Big)}{\lambda_{l}-\lambda_{m}}.
\end{eqnarray}
This result extends to skew-Hermitian matrices, $i$A, $i$B  as well. In the above formula, we can substitute A $\rightarrow$ $-i \Delta t \hspace{0.1cm} H_{u}$ and $x$B $\rightarrow$ $-i \Delta t \hspace{0.1cm} u \hspace{0.1cm} H_{1}$ with $| \lambda_{k}\rangle$ being the eigenbasis of  $A = -i \Delta t \hspace{0.1cm} H_{u} $ and $\lambda_{v} = -i \Delta t \lambda_{k}$ for $k = l, m$, where $\lambda_{k}$'s are the eigenvalues of A.

\end{document}